\documentclass[journal]{IEEEtran}
\usepackage{cite}

\ifCLASSINFOpdf
  \usepackage[pdftex]{graphicx}
  \usepackage{pgfplots}
  \usepackage{tikz}
  \graphicspath{{../pdf/}{../jpeg/}}
  \DeclareGraphicsExtensions{.pdf,.jpeg,.png}
\else
\fi
\usepackage{amsmath}
\usepackage{amssymb}
\usepackage{todonotes}
\begin{document}
%
\title{On the Impact of Fixed Point Hardware for Optical Fiber Nonlinearity Compensation Algorithms}
%
%
%

\author{Tom~Sherborne\textsuperscript{*},
        Benjamin~Banks\textsuperscript{*},
        Daniel~Semrau,~\IEEEmembership{Student~Member,~IEEE},
        Robert~I.~Killey~\IEEEmembership{Senior~Member,~IEEE},
        Polina~Bayvel,~\IEEEmembership{Fellow,~IEEE},
        and~Domani\c{c}~Lavery,~\IEEEmembership{Member,~IEEE}%

\thanks{Manuscript received January, 2018.
This work was funded by United Kingdom (UK) Engineering and Physical Sciences Research Council (EPSRC) Programme Grant UNLOC (UNLocking the capacity of Optical Communications), EP/J017582/1. D.~Lavery is supported by the Royal Academy of Engineering under the Research Fellowships scheme.}
\thanks{\textsuperscript{*}These authors contributed equally to this work. T. Sherborne and B. Banks were students within the Department
of Electronic and Electrical Engineering, University College London, London WC1E 7JE, U.K. (e-mail: zceetrs@ucl.ac.uk; zceebjb@ucl.ac.uk).}
\thanks{D. Lavery, D. Semrau and P. Bayvel are with the Optical Networks Group, Department of Electronic and Electrical Engineering, University College London, London WC1E 7JE, U.K. (e-mail: d.lavery@ucl.ac.uk; d.semrau@ucl.ac.uk; p.bayvel@ucl.ac.uk).}}


%
%

\markboth{ PREPRINT }%
{ PREPRINT }
%



\maketitle

\begin{abstract}
Nonlinearity mitigation using digital signal processing has been shown to increase the achievable data rates of optical fiber transmission links. One especially effective technique is digital back propagation (DBP), an algorithm capable of simultaneously compensating for linear and nonlinear channel distortions. The most significant barrier to implementing this technique, however, is its high computational complexity. In recent years, there have been several proposed alternatives to DBP with reduced computational complexity, although such techniques have not demonstrated performance benefits commensurate with the complexity of implementation.
In order to fully characterize the computational requirements of DBP, there is a need to model the algorithm behavior when constrained to the logic used in a digital coherent receiver. Such a model allows for the analysis of any signal recovery algorithm in terms of true hardware complexity which, crucially, includes the bit-depth of the multiplication operation. With a limited bit depth, there is  quantization noise,  introduced with each arithmetic operation, and it can no longer be assumed  that the conventional DBP algorithm will outperform its low complexity alternatives. 
In this work, DBP and a single nonlinear step DBP implementation, the \textit{Enhanced Split Step Fourier} method (ESSFM), were  compared with linear equalization using a generic software model of fixed point hardware. The requirements of bit depth and fast Fourier transform (FFT) size are discussed to examine the optimal operating regimes for these two schemes of digital nonlinearity compensation. For a 1000~km transmission system, it was found that (assuming an optimized FFT size), in terms of SNR, the ESSFM algorithm outperformed the conventional DBP for all hardware resolutions up to 13~bits. 
\end{abstract}


%
\IEEEpeerreviewmaketitle

\section{Introduction}

\IEEEpubidadjcol

\begin{figure}[t!]
   \centering
       \includegraphics[page=1,width=0.5\textwidth]{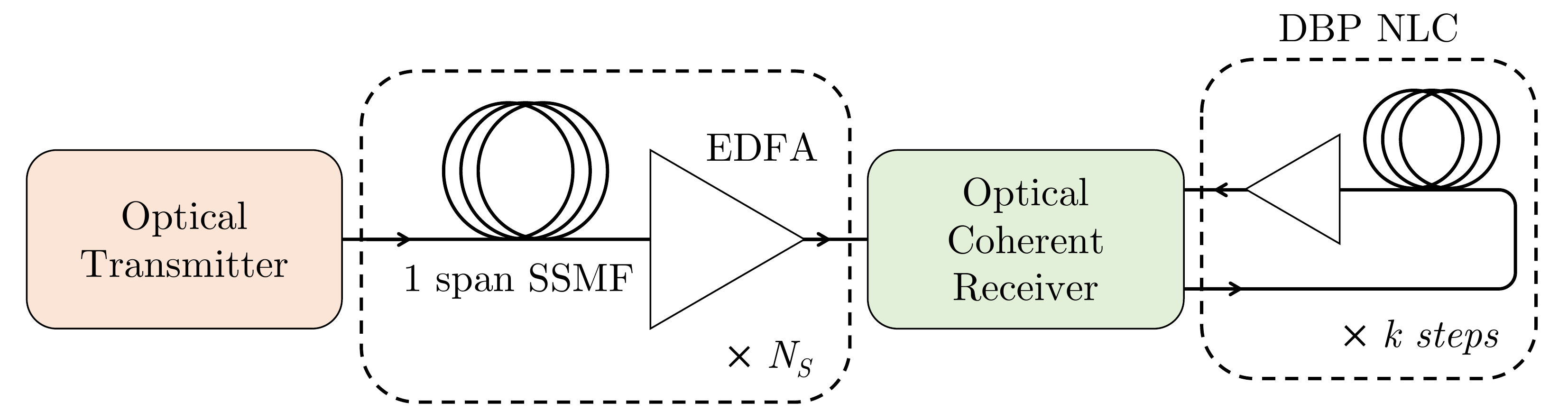} 
 \caption{Transmission model used to investigate the performance of various DBP schemes. This is implemented as a series of linear and nonlinear operators applied to the buffered signal vector. \textit{Ideal} DBP describes the most intensive implementation of DBP with $k$ steps equal to the number of steps applied over $N_{s}$ spans in simulated forward propagation.}
 \label{intro:system}
\end{figure}

\IEEEPARstart{S}{ignal} processing techniques to overcome optical fiber nonlinearity have evolved considerably in recent years. With the aim of increasing the capacity and reach of coherent detection systems, the use of digital signal processing (DSP) algorithms has been shown to significantly improve system performance \cite{Bayvel2013163} and reduce complexity \cite{NapoliPaper} when compared to earlier systems \cite{Ip2008}. 
The DBP algorithm is conventionally applied at the receiver, however it can also be applied at the transmitter (i.e., digital pre-compensation \cite{Temprana2015}), or as some combination of both link endpoints, as investigated in \cite{Lavery2016}. In all cases, the aim of DBP is to simulate a reverse `virtual' link (see Fig. \ref{intro:system}) and, consequently, reverse the deterministic nonlinear and linear effects of fiber propagation\cite{Ip2008}. 

However, the promising performance gains achieved with DBP typically incur a computational expense beyond the capabilities of present digital signal processing hardware\cite{DSMillarPaper}, due to the use of the split step Fourier method to solve the reverse Nonlinear Schr\"{o}dinger equation (NLSE) as a Manakov system\cite{AgrawalNL}. 
In order to achieve real-time processing, several implementations of DBP have been proposed with differing perspectives on the trade-off between performance and complexity. These included increasing the nonlinear step size of the split step Fourier method in the DBP virtual link to minimize the number of calculations required for signal recovery. 1~step-per-span DBP (1sps-DBP) \cite{Ip2008} has been proposed as a natural trade-off between performance and complexity for single channel DBP. More advanced proposals include the low pass filter DBP (LPF-DBP) - an implementation of DBP that introduces a phase noise filter in the nonlinear step of the algorithm\cite{Du2010}. High frequency phase noise components are present in the phase spectrum when measured across multiple spans, and the use of LPF-DBP acts to suppress such components to achieve reasonable performance with $<1$ step per span. LPF-DBP has been shown to perform comparably to 1sps-DBP with 1 step every 4~spans \cite{Du2010}, corresponding to an approximately 75\% reduction in the number of calculations required for NLC. Note that the overall numerical complexity of the aforementioned NLC schemes is dependent on the link length.

DBP is typically implemented as a Wiener-Hammerstein system\cite{DSMillarPaper} on a `virtual' link expressing a reverse journey to the physical forward propagation. The conventional Manakov system for modeling propagation concatenates operators of fixed distance to describe the full link, and so the complexity of standard DBP, which uses this approach, is proportional to the link length. A recent development in the field has been the design of NLC methods with complexity that is decoupled from the physical link length\cite{SecondiniECOC}.

For realizable NLC in a coherent receiver, any solution would ideally demonstrate performance improvement over electronic dispersion compensation (EDC) while avoiding a significant increase in computational complexity. This has spurred interest in single step-per-link implementations of DBP, which reduce the computational complexity compared to multiple step (e.g 1~step-per-span) schemes. The Enhanced Split Step Fourier Method (ESSFM) is a general implementation of filtered DBP, which applies the reversed channel with the step size equal to the link length \cite{SecondiniECOC}. Similarly to LPF-DBP, the ESSFM uses an additional operation stage in the nonlinear algorithm sub-step. The ESSFM algorithm has demonstrated a performance increase of 0.7~dB \cite{Secondini2016} over a linear Feed-Forward Equalizer (FFE) system with a comparable demand of complex multiplications\cite{Secondini2016}. Both methods of augmented DBP use nonlinear optimization routines for the determination of a nonlinear filter\cite{GaoLPF,Secondini2016}.  This routine solves a numerical problem and not the underlying problem of solving an indeterminate equation describing a physical system, and therefore it is uncertain if either method demonstrates optimality in their solutions.

These implementations of DBP have demonstrated the potential performance benefits of using NLC over EDC in offline implementations. However, to make a case for the practical deployment of such systems there remains a question of DBP performance when constrained to the fixed point arithmetic required in a high throughput coherent receiver.

This has previously been investigated in \cite{Fougstedt}, however the study was constrained by the use of time domain dispersion filters, with a manually optimized dispersion compensation ratio between the linear filters, and a specific target ASIC implementation. To extend the general understanding of quantization effects in NLC methods, here we investigate frequency domain dispersion compensation, and make no assumptions on the target hardware. A further advantage of this hardware- and system-agnostic approach is that it facilitates a clearer comparison between different NLC methods, where a change in dispersive block length can have a significant impact on performance, while removing the consideration of distortions introduced by finite-length time domain filters\cite{SavoryDigital}.

In this paper, we develop a model of finite precision arithmetic and simulate the operation of several implementations of DBP. We assess and compare the performance of constrained DBP using the linear EDC algorithm as a benchmark. Finally we determine operating regimes for which a specific implementation of NLC outperforms EDC in terms of SNR.

This paper is organized as follows. In Section \ref{section:algo_design}, different NLC algorithms are introduced. In Section \ref{section:hw_model}, the model of fixed-point arithmetic and hardware simulation is described in detail. 
The results of the transmission simulations are given in Section~\ref{section:dbp_results} for the DBP fixed point algorithm and Section~\ref{section:essfm_results} for the ESSFM fixed point algorithm. Conclusions are drawn in Section \ref{section:conc}.
\section{Methodology}
\subsection{Algorithm Design}
\label{section:algo_design}
In modeling the behavior of digital NLC we used a simulation model of an optical link (Fig.~\ref{intro:system}), assuming ideal noise-free transceivers, with the link parameters given in Table~\ref{table:params}. The signals under test were single channel 32~GBd dual polarization (DP) quadrature phase shift keying (QPSK) and 16-\textit{ary} quadrature amplitude modulation (DP-16QAM) signals. The signal was sampled at 4~samples/symbol and shaped using a root-raise cosine (RRC) filter. In this approach, we numerically solved the Manakov system for forward pulse propagation to simulate 1000~km (25$\times$40~km span) of standard single-mode fiber (SSMF), as a representative long-haul transmission system, and following the methodology in \cite{Secondini2016}.  An EDFA was included after each span to fully compensate for the power loss due to signal attenuation over the span.

\begin{table}[!t]
\caption{Summary of System Parameters}
\label{table:params}
\centering
\begin{tabular}{l|r|l}
\hline
Parameter & Value & Units\\
\hline
Fiber attenuation & 0.2 & dB/km \\
Dispersion parameter & 17 & ps/(nm $\cdot$ km)\\
Fiber nonlinear coefficient & 1.2 & 1/(W $\cdot$ km)\\
Span length & 40 & km\\
Simulation step size & 100 & m\\
1sps-DBP step size & 40,000 & m\\
1spl-DBP/ESSFM step size & Varies & m\\
Symbol rate & 32 & GBd\\
EDFA noise figure & 5 & dB\\
Pulse shape & RRC, 1\%& rolloff\\
DBP Wiener-Hammerstein split & 0.85 & - \\
ESSFM Wiener-Hammerstein split & 0.4 & - \\
\hline
\end{tabular}
\end{table}

\begin{figure}[t!]
   \centering
       \includegraphics[page=1,width=0.5\textwidth]{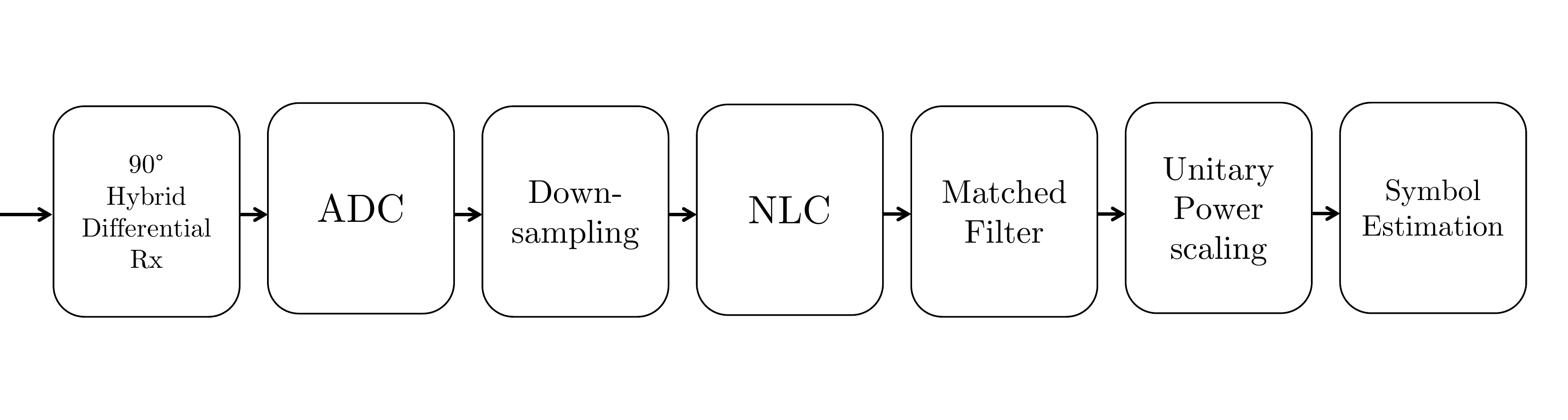} 
 \caption{
	 The signal processing chain used in this work. The complex valued pulse train is captured using a model of balanced detection, downsampled to 2 samples/symbol, NLC applied and match filtered before unitary scaling and an SNR for the symbols estimated. This routine is also the function, $f(x)$, for the optimisation of the ESSFM filter coefficients.}
 \label{algo:rx}
\end{figure}

Fig.~\ref{algo:rx} shows the receiver model used in this work. The received data is detected, resampled to 2~samples/symbol and normalized to unit average power. The signal is then processed using either EDC (implemented as a single step frequency domain filter), or the NLC algorithm under test. The approaches to NLC in this work are the ESSFM\cite{SecondiniECOC}, 1~step-per-span DBP, and 1~step-per-link DBP. In the implementation of each NLC algorithm, fixed-point arithmetic is used to model the finite precision available in practical hardware. Details of the fixed point modules are discussed in Section \ref{method:hw}. For both DBP and EDC, the Fast Fourier Transform (FFT) is based on the radix-2 Cooley-Tukey algorithm \cite{CooleyTukey} using an `overlap and save' procedure to model a buffered application of dispersion compensation. Exponentials were approximated using the coordinate rotation digital computer (CORDIC) algorithm to reduce the computational demand of approximation using a long Taylor series. The filter coefficients for the ESSFM algorithm were computed offline using a double floating-point precision non-linear optimization algorithm, but quantized to the appropriate bit depth at runtime. Following NLC or EDC, matched filtering was applied to the signal, before normalization and downsampling to 1~sample/symbol. SNR estimation was performed on the central $2^{14}$ symbols according to the method detailed in \cite{Alvarado2015} and is reported for the optimum launch power unless otherwise specified.
 
DBP was applied using the split-step Fourier method (SSFM) to solve the Manakov equation \cite{Marcuse1997} with the inverse loss, dispersion and nonlinearity fiber coefficients according to the parameters in Table \ref{table:params}. The nonlinear interaction between accumulated ASE noise and the co-propagating signal limits the channel performance even with application of \textit{ideal} DBP. This can be attributed to nonlinear phase noise accumulation (Gordon-Mollonauer effect)\cite{DSMillarPaper}. The single channel transmission at 32~GBd with 40~km spans was chosen for equitable comparison of algorithm performance between DBP and ESSFM, again, following the methodology of \cite{Secondini2016}. The Manakov inverse channel model\cite{Marcuse1997}, over some incremental step, $h$, approximates the deterministic effects of the channel as

\begin{equation}
\begin{split}
	\mathbf{E}(z+h,T) &\simeq \textrm{exp}(h\hat{D}) \textrm{exp}(h\hat{N}) \mathbf{E}(z,T) \\
	&\simeq \textrm{exp}\left(\frac{h}{2}\hat{D}_{1}\right)\textrm{exp}(L_{\textrm{eff}}\hat{N})\textrm{exp}\left(\frac{h}{2}\hat{D}_{2}\right)\mathbf{E}(z,T),
\end{split}
\label{eqtn:ssfm}
\end{equation}

where $\hat{D}$ is the linear operator, given by
\begin{equation}
	\hat{D} = \frac{i\beta_{2}}{2} \frac{\partial^2}{\partial t^2},
	\label{eqtn:marcuseD}
\end{equation}
and $\hat{N}$ is the nonlinear operator, defined as
\begin{equation}
	\hat{N} = -i\gamma\frac{8}{9}\mathbf{E^{H}E}-\frac{\alpha}{2}.
	\label{eqtn:marcusegamma}
\end{equation}
Here, $\mathbf{E} = [E_{X},E_{Y}]$ is the optical field in the $X$ and $Y$ polarization states, $\alpha$ is the fiber loss parameter, $\beta_2$ is the group velocity dispersion, $\gamma$ is the fiber nonlinearity coefficient, and $L_{\textrm{eff}}$ is the effective length of $h$.

Following the approach described in~\cite{DSMillarPaper}, a 3- block Wiener-Hammerstein model for each step was used for the DBP. Linear compensation sub-steps were performed in the frequency domain with a circular convolution using 4 FFTs per step. For DBP, adjacent linear blocks were combined over the total link for $N_{s}+1$ total FFT blocks. Nonlinear substeps were applied in the time domain as a phase shift proportional to the instantaneous power of the back propagating signal. The responsibility of application of the linear components between each $\hat{D}_{i}$ block in (\ref{eqtn:ssfm}) is dictated by the Wiener-Hammerstein (WH) split. This parameter has the effect of controlling the position of the nonlinear $\hat{N}$ term across step distance, $h$. 

To implement the ESSFM algorithm proposed in \cite{SecondiniECOC}, we use a modified nonlinear operator
\begin{equation}
\hat{N}_k = -i\gamma L_{eff}\left(\sum_{i=0}^{N_{c}} c_{i}\left(|\mathbf{E_{k-i}}|^2+|\mathbf{E_{k+i}}|^2\right)\right)
\label{eqtn:essfm-step}
\end{equation}
for each sample, $k$, with the application of the ESSFM filter coefficients, $c_i$, for filter size $N_{c}+1$. The filter coefficients are determined through a nonlinear multivariate equation solver. The DSP chain for signal processing, including the ESSFM, is formulated as some function, $f(x)$, which returns a final SNR. In this work, a Quasi-Newton cubic line search procedure was used to minimize a negated SNR output of $f(x)$. This blind optimization approach uses gradient descent with a relative tolerance of $10^{-3}$ to converge on a set of filter coefficients. Similarly to \cite{Secondini2016}, we applied nonlinear optimization to maximize the SNR of a sample sequence using the coefficients, $c_i$ as the degrees of freedom. The number of coefficients in the filter was restricted to a power of two, again for direct comparison with \cite{Secondini2016}. The number of multiplications per transmitted symbol for the ESSFM $\hat{N}$ operator can be approximated by $N_{s}(2N_{c} + 1)$ when the multiplications for instantaneous power are disregarded\footnote{Note that the ESSFM filter is a symmetric FIR filter and, thus, this filter could be efficiently implemented using frequency domain convolution. Given the relatively short filter length, the computational complexity in this case would be comparable, but this observation is noted here for completeness.}. This linear relationship suggests  the possibility for an optimality trade-off between ESSFM filter size and performance in SNR. Diminishing returns for the system in \cite{Secondini2016} were observed for filters of 32 tap weights and above. Here, this investigation was repeated for our system parameters, as it can be seen that additional multiplications in the $\hat{N}$ operator  have implications for the performance limitations from the quantization overhead in the fixed point model; there is a trade off between nonlinear performance gain and quantization noise introduced from the additional filter taps. The theoretical basis for the operation of the ESSFM and the filter coefficients is further discussed in \cite{SecondiniXPM,Secondini2016}.

As the ESSFM operates at the 1 step-per-link level, the Wiener-Hammerstein (WH) split now determines the position of the only nonlinearity compensation operator across all spans.  The optimal WH split value, previously determined to be 0.85 for conventional DBP, assumes a step, $h$, of maximum size of a single span\cite{DSMillarPaper}. We found this value for the split to be sub optimal in our system for DBP, however the SNR benefits from additional optimization only reached a maximum of $0.01$~dB and, therefore, a WH-split value of 0.85 was retained when simulating DBP as the NLC under test. However for the ESSFM, the WH split was fully re-optimized. The intuition for this is that since the distribution of power over distance for a step size of 1 span length is only a subset of the distribution over the full link, the location of the single $\hat{N}$ would likely change. In \cite{Secondini2016} the ESSFM is proposed using the Wiener Model of the $\hat{N}$ term preceding a singular $\hat{D}$ term. We found this arrangement to be sub-optimal in terms of nonlinear compensation performance, despite the additional quantization from two FFT pairs per link, and we subsequently used an optimized WH split value of 0.4.
\subsection{Hardware Model}
\label{section:hw_model}
\label{method:hw}
The primary aim of this work is to quantify the performance of different NLC algorithms when constrained to fixed-point (FXP) arithmetic. In digital circuits, signals are stored using a bit-wise representation. For $b$ bits it is possible to represent any integer value in the range $\left[0, 2^b -1\right]$. The precision of computations improves with bit depth, however high precision comes at a cost of more associated digital circuitry, and therefore complexity, which introduces greater power consumption, circuit area and, ultimately, cost. At the same time there is a need, especially in the context of a receiver, to maintain enough numerical precision so that the received signal meets a given SNR requirement. This trade-off is investigated in this work where solution precision will have a direct influence the signal-to-noise ratio (SNR) of a signal which has gone through the arithmetic process of NLC. For the different NLC schemes, it is possible to compare the performance and complexity merits of such algorithms in a realistic scenario. To do this, it is necessary to model the bit-level arithmetic occurring in realtime receiver hardware.

A software model of FXP arithmetic for an arbitrary quantization level was developed. This model remains agnostic to the hardware platform, with no intended target system. The basis for this model are functional blocks for FXP addition and multiplication, which receive some quantization level as an input parameter and return the arithmetic result. The constraints investigated here are limited to bit depth (the number of logical bits permitted in FXP arithmetic) as a primary factor in the degree of accumulated quantization and the size of the FFT window. Intuitively, a decrease in bit depth leads to greater quantization noise as the bit stream is increasingly compressed. This limits the ability of the FXP model to approximate the best case (64-bit floating point) simulation model, resulting in some level of performance degradation. Arithmetic precision is, therefore, a key design parameter which has a strong influence on the performance of any NLC scheme.

These atomic, arithmetic functions were used to emulate the operation of more complex operations, including an FFT. A fixed bit depth was assumed throughout the system and bit depth expansion was not permitted across multiple functions, as is typical in real-time signal processing hardware. However bit depth expansion is permitted for intermediate operations only\footnote{We refer to an intermediate operation as one inside a single function block but between the atomic function units. The primary usage of this intermediate bit expansion is during the FFT block, wherein conditional scaling of bit depth is employed to minimize the addition of quantization noise from successive FXP operations. This problem could also have been approached by choosing a different quantization interval, however in this work the bounds $x\in[-1,1)$ are strictly enforced.}. Each number is represented in simulation using the fractional two's complement fixed point format, with each representable value in the range $-1 \leq x < 1$. This format was chosen as there is a guarantee that the product of every multiplication will have some magnitude equal or less than that of either input. The sampling of a continuous signal to some discrete amplitude introduces quantization noise, resulting in error propagation as a result of the finite precision of the system. As a sample propagates through a cascade of multiplication stages (e.g., in the FFT) then each multiplication introduces some quantization noise. The signal-to-quantization noise ratio of a signal quantized to $B$ bits \cite{Proakis} can be described as
\begin{equation}
	SQNR = \frac{1}{\sigma ^{2}},
	\label{sqnr_eqn}
\end{equation}
where the quantization noise variance is given by
\begin{equation}
	\sigma ^{2} = \frac{2^{-2B}}{12}
\end{equation}
The logarithm of Eq.~\eqref{sqnr_eqn} reveals a gain of 6~dB in SQNR for the provision of an extra bit. The crux of Eq.~\eqref{sqnr_eqn} in the context of digital NLC is the trade-off between maximizing the number of DBP steps in order to ensure solution accuracy and minimizing the number of steps to prevent the accumulation of quantization noise.

When an input signal is first quantized to a set bit depth, each amplitude is rounded towards the nearest available discrete amplitude. In the event of an arithmetic overflow\footnote{Arithmetic overflow refers to some discrete number outside of the $[-1,1)$ bounds.} in this system, the erroneous value is clipped at maximum or minimum quantized amplitude. This method also ensures the quantization noise distribution remains zero-centered.

As previously noted, linear filtering stages of NLC are performed in the frequency domain. Transformations between the time and frequency domain are performed by a radix-2 Cooley-Tukey FFT \cite{CooleyTukey}. This format of FFT is chosen for the ability to implement a wide range of FFT sizes and because the complexity of this FFT is well studied\cite{Secondini2016}. We did not consider split radix FFTs in this work, which may have changed the overall NLC performance. However, because the same FFT implementation is common to all algorithms considered, any performance change would affect all NLC implementations. The quantization noise introduced from one butterfly operation in an FFT is $4\sigma ^{2}$ with a quantization noise variance introduced from an FFT as $4(N-1)\sigma ^{2}$\cite{Proakis}. This corresponds to the optimal SNR achievable for a signal processed using an FFT and, as such, places a ceiling on total system performance. 

For an accurate representation of online processing of a continuous stream of input samples, the incoming signal is buffered and processed in batches equal to the size of the FFT under test. The Overlap and Save algorithm was used to be able to model the continuous filtering required in hardware. The size of the buffer to process must correspond to a radix-2 FFT size, or $2^{n}$, where $n\in{}\lbrace5,\ldots15\rbrace$. An overlap size of $N/4$ was chosen, where $N$ is the length of a signal buffer, in correspondence with the work carried out in \cite{Secondini2016}. The Overlap and Save algorithm used here also applies the dispersion map as the linear substeps of DBP and the full CDC operation. Nonlinearity compensation is provided in the time domain by multiplying the signal with a vector of complex exponentials. Using Euler's identity, it is possible to approximate the complex exponential using the CORDIC algorithm\cite{KhanDSP}. In this algorithm, the sine and cosine of a provided angle is approximated iteratively, with the approximate solution accuracy improving with each iteration. Note that, in contrast to the Taylor expansion model used in \cite{Fougstedt}, CORDIC provides a greater degree of accuracy with an acceptable computational complexity.
\section{Results}
\subsection{Digital Back Propagation}
\label{section:dbp_results}
The initial simulations of the  transmission of $2^{18}$ DP-QPSK modulated samples at 32~GBd over a 1000~km fiber link. DBP was implemented with the parameters described in Section~\ref{section:algo_design} with the finite precision logic described in Section \ref{section:hw_model}. For comparison, we also implemented  CDC using the same FXP modeling constraints. The FFT size was optimized in each case and is therefore not included as a parameter. In the following, we examine the behavior of DBP in the FXP environment, and discuss the additional complexity of this form of NLC compared to CDC. 

Fig.~\ref{snrBitsDBP} shows the SNR improvement relative to EDC at the same FXP bit depth, across a range of numbers of nonlinear steps per link. For an increased number of steps in DBP, the virtual link model improves in efficacy, as expected, however this is at the cost of an increased FXP bit depth, which is required to reduce the impact of quantization noise when multiple steps are used. Additionally, Fig. \ref{snrBitsDBP} identifies that there exists some quantization limit in performance, at a given number of steps-per-link, which can be reached with a sufficient bit depth in the digital logic. In the single step per link case, with the fewest multiplications, there is a negligible SNR gain of 0.1~dB over CDC for bit depths greater than 13~bits. We observe no gain in SNR over CDC for bit depths below this, for any number of steps-per-link. The greatest performance difference over CDC is a 3.5~dB gain, observed using 50 steps per link, equivalent to 2 steps per span, was at a bit depth of 15~bits.
However, when compared to CDC it can be seen that any gain requires a minimum of 13~bits.
\begin{center}
	\begin{figure}[!tb]
		\centering
		\begin{tikzpicture}
		\begin{axis}[
		xlabel = {Bit depth},
		ylabel = {$\Delta$SNR over CDC (dB)},
		height=8cm, 
		width=0.5\textwidth, 
		grid=both,
		trim axis left,
		trim axis right,
		minor tick num=1,
		xmin = 7,
		xmax = 16,
		ymin = -5,
		ymax = 4,
		legend pos = north west,
		every axis plot/.append style={thick}] 
		
		\addplot[blue,mark = asterisk,mark size=2pt]  table[x index=0,y index=1,col sep=comma] {data/dbpOverCDC.txt};
			\addlegendentry{1 SPL}
		
		\addplot[red,mark = asterisk,mark size=2pt]  table[x index=0,y index=2,col sep=comma] {data/dbpOverCDC.txt};
			\addlegendentry{10 SPL}
		
		\addplot[orange,mark = asterisk,mark size=2pt]  table[x index=0,y index=3,col sep=comma] {data/dbpOverCDC.txt};
			\addlegendentry{25 SPL}
		
		\addplot[purple,mark = asterisk,mark size=2pt]  table[x index=0,y index=4,col sep=comma] {data/dbpOverCDC.txt};
			\addlegendentry{50 SPL}
		\end{axis}
		\end{tikzpicture}
		\caption{
		SNR improvement for DBP compared to CDC at 1000~km for DP-QPSK transmission. Both DBP and CDC are simulated with FXP arithmetic and optimized FFT size. DBP is simulated with a range of steps-per-link (SPL) to investigate the influence of the number of steps on the quantization error on the signal.}		\label{snrBitsDBP}
	\end{figure}
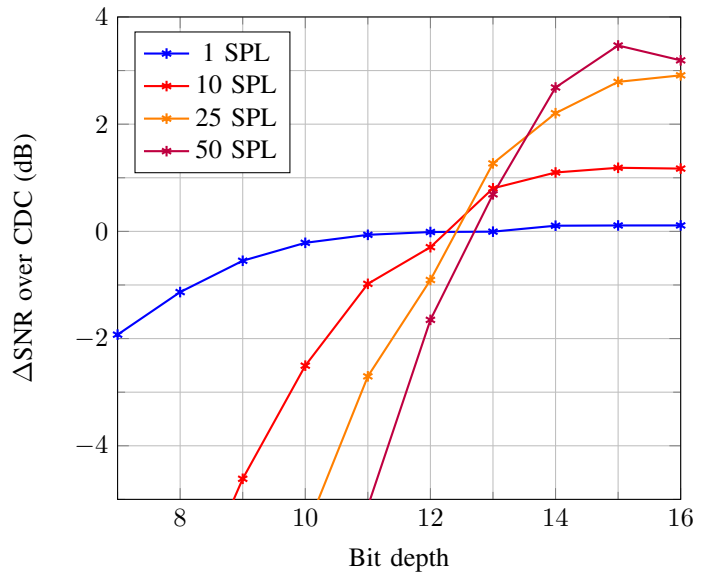
\end{center}

We expand on this analysis in Fig. \ref{splPlot} by examining the absolute SNR performance of DBP against the variation in steps-per-link between 7 and 16 bits. From this result it can be inferred that for $\leq 12$ bits, there is insufficient arithmetic precision to tolerate the increased accumulation of quantization noise arising from increasing the number of DBP steps per link. While increasing the steps per link increases the accuracy of theoretical DBP our results show that increasing the number of FFTs and multiplications is monotonically detrimental for low bit depths due to quantization error. However for $\geq 13$~bits, there is a noted transition to an operating regime for which an increase in the steps per link results in some SNR gain. This confirms that at 1000~km, a minimum of 13~bits are required to overcome the quantization overhead present in our implementation of DBP. In particular for 13~bits, there is an increase in SNR as a result of increased steps up to 25~SPL, where a peak SNR of $23.6$~dB is observed. Beyond this a subsequent decrease in SNR is observed, likely as the quantization at 50~SPL becomes dominant even for 13~bits resolution. Another way to view this result is that there exists an \textit{optimum} number of steps per link for a given FXP hardware bit depth. Previously reported results in this area show that DBP performance increases monotonically with number of steps per link and, thus, this result highlights the importance of considering fixed point arithmetic in performance analyses.

\begin{center}
\begin{figure}[!tb]
	\centering
	\begin{tikzpicture}
	\begin{semilogxaxis}[
	xlabel = {Steps per link},
	ylabel = {SNR (dB)},
	height=8cm, 
	width=0.45 \textwidth, 
	grid=both,
	trim axis left,
	trim axis right,
	minor tick num=1,
	xmin = 1,
	xmax = 50,
	ymin = -5,
	ymax = 27.5,
    	xticklabels={$1$,$ $,$10$,$ $,$50$},
	legend pos = south west,
	every axis plot/.append style={thick}] 
	\addplot[blue,mark = asterisk,mark size=2pt]  table[x index=0,y index=1,col sep=comma] {data/splData.txt}; 
	\addlegendentry{7 bits}
	\addplot[red,mark = asterisk,mark size=2pt]  table[x index=0,y index=2,col sep=comma] {data/splData.txt}; 
	\addlegendentry{8 bits}
	\addplot[orange,mark = asterisk,mark size=2pt]  table[x index=0,y index=3,col sep=comma] {data/splData.txt}; 
	\addlegendentry{9 bits}
	\addplot[purple,mark = asterisk,mark size=2pt]  table[x index=0,y index=4,col sep=comma] {data/splData.txt}; 
	\addlegendentry{10 bits}
		\addplot[green!50!blue,mark = asterisk,mark size=2pt]  table[x index=0,y index=5,col sep=comma] {data/splData.txt}; 
	\addlegendentry{11 bits}
		\addplot[blue,mark = diamond,mark size=2pt]  table[x index=0,y index=6,col sep=comma] {data/splData.txt}; 
	\addlegendentry{12 bits}
		\addplot[red,mark = diamond,mark size=2pt]  table[x index=0,y index=7,col sep=comma] {data/splData.txt}; 
	\addlegendentry{13 bits}
			\addplot[orange,mark = diamond,mark size=2pt]  table[x index=0,y index=8,col sep=comma] {data/splData.txt}; 
	\addlegendentry{14 bits}
			\addplot[purple,mark = diamond,mark size=2pt]  table[x index=0,y index=9,col sep=comma] {data/splData.txt}; 
	\addlegendentry{15 bits}
				\addplot[green!50!blue,mark = diamond,mark size=2pt]  table[x index=0,y index=10,col sep=comma] {data/splData.txt}; 
	\addlegendentry{16 bits}

	\end{semilogxaxis}
	\end{tikzpicture}
	\caption{
	SNR against the steps per link for DBP at 1000~km for 32~GBd DP-QPSK transmission. DBP is simulated with FXP arithmetic with the number of steps in the virtual fiber model and bit depth varied. 
	}
	\label{splPlot}
\end{figure}
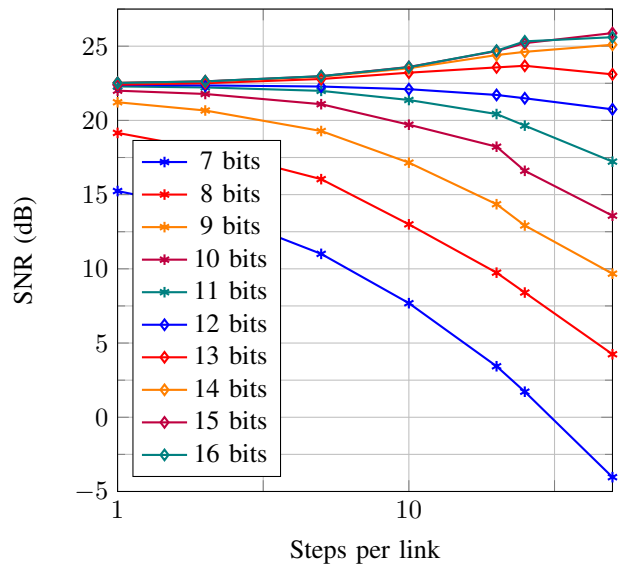
\end{center}
\subsection{Enhanced Split Step Fourier Method}
\label{section:essfm_results}

	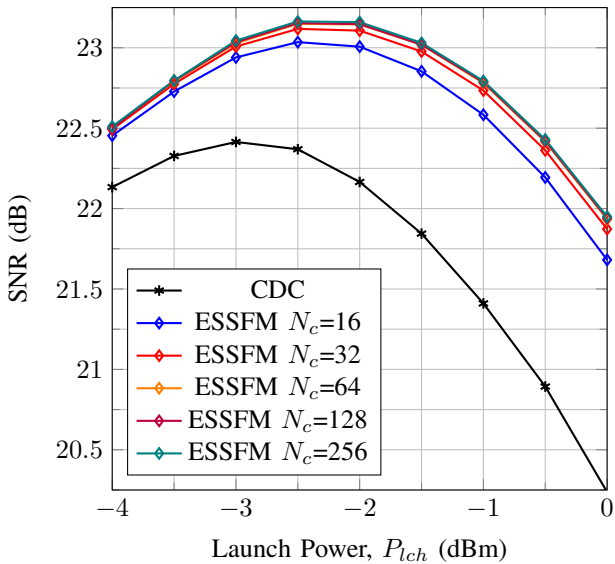
\begin{figure}[!b]
		\centering
		\begin{tikzpicture}
		\begin{axis}[
		xlabel = {Launch Power, $P_{lch}$ (dBm)},
		ylabel = {SNR (dB)},
		height=8cm, 
		width=0.45\textwidth, 
		grid=both,
		trim axis left,
		trim axis right,
		minor tick num=1,
		xmin = -4,
		xmax = 0,
		ymin = 20.25,
		ymax = 23.25,
		legend pos = south west,
		every axis plot/.append style={thick}] 
\addplot[black,mark = asterisk,mark size=2pt]  table[x index=0,y index=1,col sep=comma] {data/essfm_4qam_1000km_snr_plch_zoom.txt};
			\addlegendentry{CDC}
		\addplot[blue,mark = diamond,mark size=2pt]  table[x index=0,y index=2,col sep=comma] {data/essfm_4qam_1000km_snr_plch_zoom.txt};
			\addlegendentry{ESSFM $N_{c}$=16}		
		\addplot[red,mark = diamond,mark size=2pt]  table[x index=0,y index=3,col sep=comma] {data/essfm_4qam_1000km_snr_plch_zoom.txt};
			\addlegendentry{ESSFM $N_{c}$=32}	
		\addplot[orange,mark = diamond,mark size=2pt]  table[x index=0,y index=4,col sep=comma] {data/essfm_4qam_1000km_snr_plch_zoom.txt};
			\addlegendentry{ESSFM $N_{c}$=64}	
		\addplot[purple,mark = diamond,mark size=2pt]  table[x index=0,y index=5,col sep=comma] {data/essfm_4qam_1000km_snr_plch_zoom.txt};
			\addlegendentry{ESSFM $N_{c}$=128}	
		\addplot[green!50!blue,mark = diamond,mark size=2pt]  table[x index=0,y index=6,col sep=comma] {data/essfm_4qam_1000km_snr_plch_zoom.txt};
			\addlegendentry{ESSFM $N_{c}$=256}		
		\end{axis}
		\end{tikzpicture}
		\caption{SNR against launch power for the ESSFM applied as NLC to a 32~GBd DP-QPSK signal. The CDC baseline is also shown. This simulation was produced using double-precision floating point arithmetic, representing the ceiling of achievable performance for a comparable fixed point simulation which will have a non-negligible quantization penalty.}
		\label{essfm:snr_plch}
	\end{figure}

The ESSFM algorithm described in Section \ref{section:algo_design} was simulated in the system detailed by Fig.~\ref{algo:rx}. For the DP-QPSK scenario, Fig.~\ref{essfm:snr_plch} identifies the variation in SNR against launch power. Each power and coefficient filter size was separately optimized up to a filter of size $N_{c}=256$. 

Improvements in SNR compared to CDC were observed for all input powers near the optimum. The maximum SNR improvement was 0.7~dB for DP-QPSK and 0.8~dB for DP-16QAM with a 256-tap filter.  It should be noted that the optimum launch power increased to -2.5~dBm for both the DP-QPSK and DP-16QAM cases. These two results together give confidence in the effective operation of the ESSFM, and are comparable to the original results for this algorithm as presented in \cite{Secondini2016}.

Similarly to previous work, marginal gains in SNR were observed for significant increases in the filter size. Fig.~\ref{essfm:dsnr_taps} highlights the aforementioned trade-off between the filter size and the efficacy of the NLC operator. This result indicates that the algorithm saturates for the given system parameters near $N_{c}=128$, demonstrating only 0.1~dB increase in SNR for a doubling of the filter size and increased hardware complexity. For the remainder of this work, the ESSFM reported in this work uses a filter of size $128$ because of this observed saturation. 

\begin{center}
	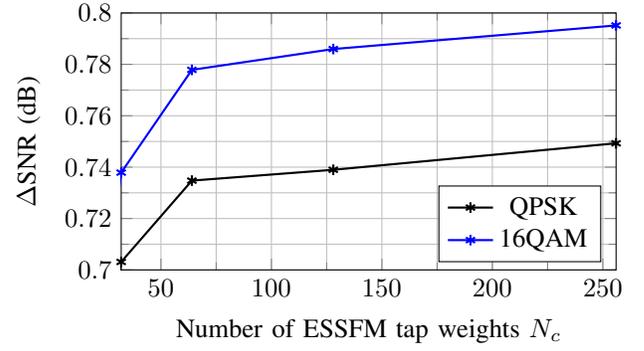
\begin{figure}
		\centering
		\begin{tikzpicture}
		\begin{axis}[
		xlabel = {Number of ESSFM tap weights $N_{c}$},
		ylabel = {$\Delta$SNR (dB)},
		height=5cm, 
		width=0.45 \textwidth, 
		grid=both,
		trim axis left,
		trim axis right,
		minor tick num=1,
		xmin = 32,
		xmax = 256,
		ymin = 0.7,
		ymax = 0.8,
		legend pos = south east,
		every axis plot/.append style={thick}] 
\addplot[black,mark = asterisk,mark size=2pt]  table[x index=0,y index=1,col sep=comma] {data/essfm_4qam_1000km_dsnr_nc.txt};
			\addlegendentry{QPSK}
\addplot[blue,mark = asterisk,mark size=2pt]  table[x index=0,y index=1,col sep=comma] {data/essfm_16qam_1000km_dsnr_nc.txt};
			\addlegendentry{16QAM}	
		\end{axis}
		\end{tikzpicture}
		\caption{$\Delta$SNR over CDC for the ESSFM at optimum $P_{lch}$ for increasing filter size.}
		\label{essfm:dsnr_taps}
	\end{figure}
\end{center}

Previous research 
into modifying the nonlinear operator in DBP\cite{Secondini2016,GaoMethodBW,GaoLPF} used a black box approach generating an optimal filter. We follow this methodology, however we chose to inspect further the behavior of the filter returned through the optimization routine detailed in Section~\ref{section:algo_design}. In Fig.~\ref{essfm:fvtool}, the 128-tap filter for processing a DP-QPSK signal at the optimum launch power is analyzed in the frequency domain. The profile of this filter is effectively a weighting of a neighboring samples' power to each sample. Fig.~\ref{essfm:fvtool} demonstrates that this filter exhibits a low pass filtering behavior that was not a constraint intended as part of the design. We note that this highlights more similarity between the methods of the LPF-DBP and the ESSFM, as the coefficients in the ESSFM were not originally proposed as a filter of any specific kind. With the understanding that the $\check{N}$ operator in the ESSFM is performing a linear phase\footnote{The linear phase response of the ESSFM 128-tap filter is not shown in this work for brevity. Also not shown is similar low pass behavior for the 128-tap filter designed for DP-16QAM.} low pass filtering with one NLC term per link. We suggest that in future work, the approaches of \cite{GaoMethodBW} and \cite{Secondini2016} could be integrated, e.g., through direct optimization of the filter bandwidth within the ESSFM.

\begin{center}
	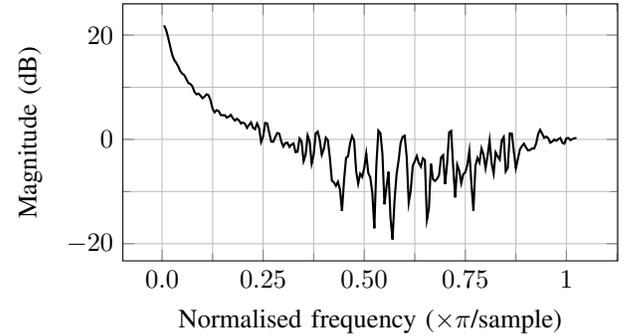
\begin{figure}[!htb]
		\centering
		\begin{tikzpicture}
		\begin{axis}[
		xlabel = {Normalised frequency ($\times\pi$/sample)},
		ylabel = {Magnitude (dB)},
		height=5cm, 
		width=0.45 \textwidth, 
		grid=both,
		trim axis left,
		trim axis right,
		minor tick num=1,
    		xticklabels={$0$,$0.0$,$0.25$,$0.50$,$0.75$,$1$},
		trim axis left,
		trim axis right,
		every axis plot/.append style={thick}] 
\addplot[black,mark size=2pt]  table[x index=0,y index=1,col sep=comma] {data/essfm_1000km_nc128_-2.5dBm_fvtool_decimate_40.txt};
		\end{axis}
		\end{tikzpicture}
		\caption{Frequency response of the 128-tap ESSFM $\check{N}$ operator for 32~GBd DP-QPSK transmission.}
		\label{essfm:fvtool}
	\end{figure}
\end{center}

Finally, the behavior of the ESSFM as NLC in a finite precision arithmetic simulation is explored. 
Fig. \ref{essfm:cut_thru} shows the maximum SNR improvement over FXP CDC with a varying number of bits and FFT size optimized for each bit depth. It can be seen that, as anticipated, there exists some quantization penalty limiting the finite precision algorithm performance when compared to double precision. The difference between the overheads of 0.024~dB for DP-QPSK and 0.18~dB for DP-16QAM highlights a potential level of increased sensitivity to quantization for multi-level modulation formats. As the level of quantization noise is constant for any distance with the ESSFM, there exists some link length at which channel noise dominates over quantization noise and this overhead becomes negligible. The exact location of this crossover point was not explored, as it is an inherently system-dependent parameter, however it is estimated from preliminary results to be near a link length of 2000~km for the system considered herein.

\begin{center}
	\begin{figure}
		\centering
		\begin{tikzpicture}
		\begin{axis}[
		xlabel = {Number of bits},
		ylabel = {$\Delta$SNR (dB)},
		height=8cm, 
		width=0.45 \textwidth, 
		grid=both,
		trim axis left,
		trim axis right,
		minor tick num=1,
		xmin = 6,
		xmax = 16,
		ymin = -3,
		ymax = 1,
		legend pos = south east,
		every axis plot/.append style={thick}] 
\addplot[black,mark = diamond,mark size=2pt]  table[x index=0,y index=1,col sep=comma] {data/essfm_cdc_cut_1000km_4_16QAM_simple.txt};
			\addlegendentry{QPSK}
\addplot[blue,mark = diamond,mark size=2pt]  table[x index=0,y index=2,col sep=comma] {data/essfm_cdc_cut_1000km_4_16QAM_simple.txt};
			\addlegendentry{16QAM}	
		\end{axis}
		\end{tikzpicture}
		\caption{$\Delta$SNR over CDC for the ESSFM against bit depth at 1000~km for 32~GBd DP-QPSK and DP-16QAM transmission.}
		\label{essfm:cut_thru}
	\end{figure}
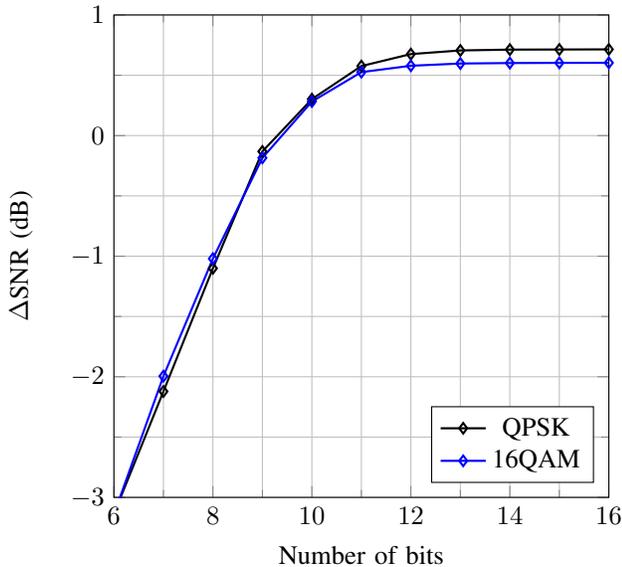
\end{center}

The results in Fig. \ref{essfm:cut_thru} also highlight the limitations of even low complexity DBP when compared to CDC, as a minimum of 10~bits are required to improve over linear equalization. To achieve gains above 0.7~dB over CDC in the DP-QPSK case, at least 13~bits are required, with 14~bits required for the DP-16QAM algorithm to exhibit a benefit of 0.6~dB in SNR. These requirements for bit depth are greater than the operating size of typical ASICs used in high speed coherent receivers, and present a barrier to achieving the full SNR gains promised by this NLC algorithm. Without further investigation into nonlinear signal recovery, our work appears to show the continued strength of linear algorithms for hardware implementation.

Comparing Fig.~\ref{snrBitsDBP} and Fig.~\ref{essfm:cut_thru}, we can observe that, at a 1~step-per-link level, the low complexity ESSFM demonstrates improvements in SNR performance when compared to both DBP and EDC. However, this gain in SNR remains less than $1$~dB and is sub-optimal when compared to a DBP using a greater number of steps-per-link. Provided there is a sufficient number of bits to saturate the system performance, Fig.~\ref{snrBitsDBP} demonstrates that more steps-per-link will yield greater SNR improvement compared to the addition of the nonlinear phase filter in the ESSFM.  Nevertheless, it should be noted that the ESSFM outperforms DBP for all bit depths $\leq$13~bits, irrespective of the number of steps used for DBP.
\section{Conclusion}
\label{section:conc}
The behavior of algorithms for optical fiber nonlinearity compensation were investigated in detail under the constraint of fixed-point arithmetic. Using a simulation including an ideal transmitter and coherent receiver, limitations on performance for signal recovery in DSP, have been shown, and  are primarily  a result of quantization noise. It was found that a minimum of 13~bits is required to observe any additional SNR gain from $\geq{}1$ steps-per-link DBP. A similar bit depth is required to reach the maximum achievable performance benefit over CDC for the ESSFM algorithm, however we note that some improvement over CDC can be seen with the reduced constraint of 10~bits. Although other algorithms for simplified DBP exist \cite{Schmauss2011, Hager2018}, the work presented herein has shown that it is crucial to include an analysis of bit depth in order to accurately compare the performance of different NLC algorithms, and that the number of mathematical operations, alone, cannot serve as a proxy for complexity.

We also highlight, for the first time, the similarities in behavior of the ESSFM and LPF-DBP through the observation that the frequency response of the ESSFM phase filter exhibits low pass filtering behavior. Future work on this topic could include the investigation into the behavior of the aforementioned NLC in a multi-channel or super-channel system, as well as expanding on the impact of modulation format on the quantization penalty of the finite-precision ESSFM.

More generally, we note that the fixed point models developed herein could be used as a basis for estimating the true performance of any NLC algorithm, and should, therefore, be included as inherent consideration in the NLC algorithm design process.

%


\section*{Acknowledgment}
The authors wish to extend their thanks to Drs Rachid Bouziane and Philip Watts, for enlightening discussions on the topic of hardware design, and assistance with developing the models used in this work.
%

\ifCLASSOPTIONcaptionsoff
  \newpage
\fi



\bibliographystyle{IEEEtran}


\begin{thebibliography}{10}
\providecommand{\url}[1]{#1}
\csname url@samestyle\endcsname
\providecommand{\newblock}{\relax}
\providecommand{\bibinfo}[2]{#2}
\providecommand{\BIBentrySTDinterwordspacing}{\spaceskip=0pt\relax}
\providecommand{\BIBentryALTinterwordstretchfactor}{4}
\providecommand{\BIBentryALTinterwordspacing}{\spaceskip=\fontdimen2\font plus
\BIBentryALTinterwordstretchfactor\fontdimen3\font minus
  \fontdimen4\font\relax}
\providecommand{\BIBforeignlanguage}[2]{{%
\expandafter\ifx\csname l@#1\endcsname\relax
\typeout{** WARNING: IEEEtran.bst: No hyphenation pattern has been}%
\typeout{** loaded for the language `#1'. Using the pattern for}%
\typeout{** the default language instead.}%
\else
\language=\csname l@#1\endcsname
\fi
#2}}
\providecommand{\BIBdecl}{\relax}
\BIBdecl

\bibitem{Bayvel2013163}
\BIBentryALTinterwordspacing
P.~Bayvel, C.~Behrens, and D.~S. Millar, ``Chapter 5 - {D}igital {S}ignal
  {P}rocessing ({DSP}) and its application in optical communication systems,''
  in \emph{Optical Fiber Telecommunications (Sixth Edition)}, 6th~ed., ser.
  Optics and Photonics.\hskip 1em plus 0.5em minus 0.4em\relax Boston: Academic
  Press, 2013, pp.~163--219.

\bibitem{NapoliPaper}
A.~Napoli, Z.~Maalej, V.~A. J.~M. Sleiffer, M.~Kuschnerov, D.~Rafique,
  E.~Timmers, B.~Spinnler, T.~Rahman, L.~D. Coelho, and N.~Hanik, ``{Reduced
  complexity digital back-propagation methods for optical communication
  systems},'' \emph{Journal of Lightwave Technology}, vol.~32, no.~7, pp.
  1351--1362, 2014.

\bibitem{Ip2008}
\BIBentryALTinterwordspacing
E.~Ip and J.~M. Kahn, ``{Compensation of dispersion and nonlinear impairments
  using digital backpropagation},'' \emph{Journal of Lightwave Technology},
  vol.~26, no.~20, pp.~3416--3425, 2008.

\bibitem{Temprana2015}
\BIBentryALTinterwordspacing
E.~Temprana, E.~Myslivets, L.~Liu, V.~Ataie, A.~Wiberg, B.~Kuo, N.~Alic, and
  S.~Radic, ``Two-fold transmission reach enhancement enabled by
  transmitter-side digital backpropagation and optical frequency comb-derived
  information carriers,'' \emph{Opt. Express}, vol.~23, no.~16, pp.
  20\,774--20\,783, Aug 2015.

\bibitem{Lavery2016}
D.~Lavery, D.~Ives, G.~Liga, A.~Alvarado, S.~J. Savory, and P.~Bayvel, ``{The
  Benefit of Split Nonlinearity Compensation for Single-Channel Optical Fiber
  Communications},'' \emph{IEEE Photonics Technology Letters}, vol.~28, no.~17,
  pp. 1803--1806, 2016.

\bibitem{DSMillarPaper}
D.~S. Millar, S.~Makovejs, C.~Behrens, S.~Hellerbrand, R.~I. Killey, P.~Bayvel,
  and S.~J. Savory, ``{Mitigation of fiber nonlinearity using a digital
  coherent receiver},'' \emph{IEEE Journal on Selected Topics in Quantum
  Electronics}, vol.~16, no.~5, pp. 1217--1226, 2010.

\bibitem{AgrawalNL}
\BIBentryALTinterwordspacing
G.~P. Agrawal, \emph{Applications of Nonlinear Fiber Optics (2nd Edition)},
  2nd~ed.\hskip 1em plus 0.5em minus 0.4em\relax Elsevier, 2008.

\bibitem{Du2010}
\BIBentryALTinterwordspacing
L.~B. Du and A.~J. Lowery, ``{Improved single channel backpropagation for
  intra-channel fiber nonlinearity compensation in long-haul optical
  communication systems.}'' \emph{Optics express}, vol.~18, no.~16, pp.
  17\,075--17\,088, 2010.

\bibitem{SecondiniECOC}
M.~Secondini, D.~Marsella, and E.~Forestieri, ``{Enhanced split-step Fourier
  method for digital backpropagation},'' \emph{European Conference on Optical
  Communication, ECOC}, no.~1, pp. 3--5, 2014.

\bibitem{Secondini2016}
M.~Secondini, S.~Rommel, G.~Meloni, F.~Fresi, E.~Forestieri, and L.~Pot{\`{i}},
  ``{Single-step digital backpropagation for nonlinearity mitigation},''
  \emph{Photonic Network Communications}, vol.~31, no.~3, pp. 493--502, 2016.

\bibitem{GaoLPF}
Y.~Gao, J.~H. Ke, J.~C. Cartledge, and S.~Yam, ``{Low-Pass Filter Assisted
  Digital Back Propagation Algorithm for {112~Gb/s} {DP 16-QAM}},'' \emph{Paper SPT5D.5 in Proc. Signal Processing in Photonic Communications (SPPCom) 2013}.

\bibitem{Fougstedt}
\BIBentryALTinterwordspacing
C.~Fougstedt, M.~Mazur, L.~Svensson, H.~eliasson, M.~Karlsson, and
  P.~Larsson-Edefors, ``Time-domain digital back propagation: Algorithm and
  finite-precision implementation aspects,'' in \emph{Optical Fiber
  Communication Conference}.\hskip 1em plus 0.5em minus 0.4em\relax Optical
  Society of America, 2017, p. W1G.4.

\bibitem{SavoryDigital}
S.~J. Savory, ``Digital filters for coherent optical receivers,'' \emph{Optics
  Express}, vol.~16, no.~2, 2008.

\bibitem{CooleyTukey}
J.~W. Cooley and J.~W. Tukey, ``An algorithm for the machine calculation of
  complex fourier series,'' \emph{Mathematics of Computation}, vol.~19, no.~90,
  1965.

\bibitem{Alvarado2015}
A.~Alvarado, E.~Agrell, D.~Lavery, R.~Maher, and P.~Bayvel, ``Replacing the
  soft-decision {FEC} limit paradigm in the design of optical communication
  systems,'' \emph{Journal of Lightwave Technology}, vol.~33, no.~20, pp.
  4338--4352, Oct 2015.

\bibitem{Marcuse1997}
D.~Marcuse, C.~R. Menyuk, and P.~K.~A. Wai, ``{Application of the Manakov-PMD
  equation to studies of signal propagation in optical fibers with randomly
  varying birefringence},'' \emph{Journal of Lightwave Technology}, vol.~15,
  no.~9, pp. 1735--1745, 1997.

\bibitem{SecondiniXPM}
M.~Secondini and E.~Forestieri, ``{On XPM mitigation in WDM fiber-optic
  systems},'' \emph{IEEE Photonics Technology Letters}, vol.~26, no.~22, pp.
  2252--2255, 2014.

\bibitem{Proakis}
J.~G. Proakis and D.~G. Manolakis, \emph{Digital Signal Processing},
  4th~ed.\hskip 1em plus 0.5em minus 0.4em\relax Pearson Education
  International, 2007.

\bibitem{KhanDSP}
S.~A. Khan, \emph{Digital Design of Signal Processing Systems: A Practical
  Approach}.\hskip 1em plus 0.5em minus 0.4em\relax John Wiley \& Sons, Ltd,
  2011.

\bibitem{GaoMethodBW}
Y.~Gao, J.~H. Ke, J.~C. Cartledge, and S.~S.~H. Yam, ``Method for determining
  the low-pass filter bandwidth for the low-pass filter assisted digital back
  propagation algorithm,'' in \emph{39th European Conference and Exhibition on
  Optical Communication (ECOC 2013)}, Sept 2013, pp. 1--3.

\bibitem{Schmauss2011}
R.~Asif, C.~Y. Lin, M.~Holtmannspoetter, and B.~Schmauss, ``Logarithmic
  step-size based digital backward propagation in {N}-channel {112Gbit/s/ch}
  {DP-QPSK} transmission,'' in \emph{2011 13th International Conference on
  Transparent Optical Networks}, June 2011, pp. 1--4.

\bibitem{Hager2018}
C.~H\"{a}ger and H.~D. Pfister, ``Nonlinear interference mitigation via deep
  neural networks,'' in \emph{Optical Fiber Communication Conference}.\hskip
  1em plus 0.5em minus 0.4em\relax Optical Society of America, 2018, p. W3A.4.

\end{thebibliography}

%
%

%
%
%




\end{document}